\documentclass[showpacs,preprintnumbers,amsmath,
amssymb,nofootinbib]{revtex4}
\usepackage{graphicx,bm}% Include figure files
\usepackage{dcolumn}    % Align table columns on decimal point
\usepackage{epsf,psfig}
\input epsf

\newcommand{\beq}{\begin{equation}}
\newcommand{\beqa}{\begin{eqnarray}}
		  \newcommand{\eeq}{\end{equation}}
\newcommand{\eeqa}{\end{eqnarray}}

\newcommand{\lsim}{\lesssim}
\newcommand{\gsim}{\gtrsim}

\newcommand{\vect}[1]{\mbox{\boldmath${#1}$}} 
\newcommand{\lmk}{\left(}
\newcommand{\rmk}{\right)}
\newcommand{\lnk}{\left\{ }
\newcommand{\rnk}{\right\} }
\newcommand{\lkk}{\left[}
\newcommand{\rkk}{\right]}
\newcommand{\lla}{\left\langle}

\newcommand{\rra}{\right\rangle}

\newcommand{\cA}{{\cal A}}

\newcommand{\vex}{{\vect x}}

\newcommand{\ven}{\hat{\vect n}}

\newcommand{\vev}{\hat {\vect v}}

\newcommand{\ved}{{\vect d}}
\newcommand{\ve}{{\vect e}}

\newcommand{\veu}{\hat {\vect u}}

\begin{document}

\title{ Non-Gaussianity analysis of GW background made by short-duration burst signals } 
\author{Naoki Seto}
\affiliation{Department of Physics, Kyoto University
Kyoto 606-8502, Japan
}

\date{\today}

\begin{abstract}
We study an observational method to analyze non-Gaussianity of a gravitational wave (GW) background made by superposition of weak burst signals. The proposed method is based on fourth-order correlations of data from four detectors, and might be useful to discriminate the origin of a GW background. 
With a formulation newly developed to discuss geometrical aspects of the correlations,
it is found that the method provides us with linear combinations of two interesting parameters, $I_2$ and $V_2$ defined by the Stokes parameters of individual GW burst  signals.  We also evaluate sensitivities of  specific detector networks to these parameters.

\end{abstract}
\maketitle

\section{introduction}

In the last decade, sensitivities of gravitational wave (GW) detectors have been dramatically improved \cite{Ando:2001ej,Acernese:2002bw,Abbott:2006zx}.   It is expected that a similar trend would continue in the next decade,  and  we will soon detect GWs directly.  Then a totally  new branch of astronomy will be opened with various observational targets from astrophysics to  fundamental physics.  One of the most interesting targets of GW observation is a stochastic GW background \cite{Allen:1996vm,Flanagan:1993ix,allen,Abbott:2006zx}.  For example, we might obtain crucial information of physics at an extremely  high-energy  scale by analyzing a GW background generated in the early universe.  Around $f\sim$100Hz,  the measurement sensitivity of the Laser Interferometer Gravitational Wave Observatory (LIGO)
to a GW  background  is about to surpass  the level $\Omega_{GW}(f)\sim 10^{-5}$ that can be indirectly constrained by the observed abundances of light elements composed at the epoch of nucleosynthesis \cite{Allen:1996vm} (see also \cite{Smith:2006nka}).  Here the function  $\Omega_{GW}(f)$ is the energy density of a GW background per logarithmic frequency interval  normalized by the critical density of the universe \cite{Allen:1996vm}.
In order to efficiently extract information from a GW background, it is essential to quantify it with various measures in addition to  the traditional spectrum $\Omega_{GW}(f)$  (see {\it e.g.} \cite{Giampieri:1997ie} for anisotropies and \cite{Seto:2006hf,Seto:2007tn} for asymmetry of right- and left-handed GWs). 

In a recent paper \cite{Seto:2008xr} (hereafter Paper 1), the author discussed detectability of non-Gaussianity of  a GW background caused by short-duration bursts whose signals are individually weak and undetectable (see also \cite{Finn:1999vh,Drasco:2002yd,Racine:2007gv,Coward:2006df}).  The origin of the non-Gaussianity here is  the discreteness and finiteness of the numbers  of the burst events.  The proposed method might be useful to discriminate the genesis of a background {\it e.g.} whether it is smooth enough to be consistent with that generated during an inflation epoch.   In addition, the author pointed out that by a statistical amplification of weak bursts signals, we might estimate their basic characters such as  rate and amplitude of bursts and their typical duration.    In this follow-up paper, we extend the previous study with fixing some overlooked  points in Paper 1, and also provide a new formulation for quantitatively discussing  the non-Gaussianity measurement in relation to the  geometry of a detector network and polarization properties of burst GW emissions.

This paper is organized as follows;   in section II, using a simplified model, we explain our basic approach for  the non-Gaussianity measurement based on fourth-order correlations, and discuss application of this approach to observational studies on  GW backgrounds.  In section III, we analyze how the  non-Gaussianity measurement depends on the geometry of  a detector network.  We define {\it the generalized overlap reduction functions} $\zeta_{II}$ and $\zeta_{VV}$ that characterize sensitivity of the measurement for a given four-detector network.  Then, in section IV, the generalized overlap reduction functions are evaluated for two specific networks of GW detectors both on the Earth and in space.  In section V, we calculate expressions for estimating signal-to-noise ratios of our non-Gaussianity measurement.  Section VI is a brief summary  of this paper.   Two appendixes are added.   In appendix A, we examine correlations of Fourier modes transformed with a short time duration.  This appendix is technical, but would be useful to understand some of basic properties  in this paper.   In appendix B, we comment on the overlooked points in Paper 1.

\section{detecting non-Gaussianity}

\subsection{underling approach}
In this paper, we mainly study a GW background made by  a superposition of burst-like signals, and discuss how to analyze its non-Gaussianity. As a preliminary set-up, we first outline the underlying approach using a simplified model.
We consider four detectors $a,b,c$, and  $d$ with representing their  data  $u_{ai},u_{bi},u_{ci}$ and $u_{di}$ respectively. Here $i$ is the suffix for the data sequence  ({\it e.g.} Fourier modes).  But, since the important issue in this subsection is the inter-detector correlation structure within a single $i$ and   not the correlation between different $i$,  we omit the suffix $i$ below.  

We assume that data $u_J$ $(J=a,b,c,d$) are superpositions of the following three ingredients; (i)  detector noises $n_J$ with no correlation between other detectors, (ii) a Gaussian signal $g$ (mimicking a smooth inflation-type GW background) and (iii) a group of { independent} burst signals $\sum_j^l v_j$.   Here $l$ is the number of bursts in the data and a  random variable.   Then the four data are written as
\beq
u_{a}=n_{a}+g+\sum_j^l v_{j},~~u_{b}=n_{b}+g+\sum_j^l v_{j},~~u_{c}=n_{c}+g+\sum_j^l v_{j},~~u_{d}=n_{d}+g+\sum_j^l v_{j}.
\eeq
 The latter two ingredients $g$ and $\sum_j^l v_{j}$ are set to be identical for all the detectors, and we dropped the label $J$ for these two components.  This is just for simplicity  and is not essential for our demonstration of extracting  non-Gaussian signature. In the next subsection, we deal with the differences between detectors taking into account their responses to incident burst signals.

 We assume that the number of bursts $l$ obeys the  Poisson statistics whose probability distribution function $P(l)$  is solely determined by the averaged event number $q$ as
\beq
P(l)=\frac{e^{-q} q^l}{l!}.
\eeq
Using this functional form, we can derive the following identities  for moments
\beq
\sum_{l=0}^\infty P(l) =1,~~~\sum_{l=0}^\infty P(l) l=q,~~~\sum_{l=0}^\infty P(l) l(l-1)=q^2.
\eeq
The first expression shows the proper normalization of the probability distribution function, while the second one is the definition of the averaged event rate $q$.

So far  we have discussed the data $u_J$ in somewhat abstract manner.
In the standard correlation analysis of a GW background, the data $u_J$ are the Fourier modes transformed from  data streams acquired in the time domain. Therefore, we treat our data $u_J$ as complex numbers, and  define moments required for evaluating correlations of the data $u_J$.
We represent  moments of the Gaussian signal $g$ as
\beq
G_1=\lla g g^*\rra,~~G_2=\lla g g\rra,\label{e5}
\eeq
where the notation $\lla \cdots \rra$ represents to take an ensemble average.
Even if the original ({\it e.g.} time domain) data stream is a real and Gaussian variable, the amplitudes of 
the real and imaginary parts of its Fourier mode $g$ can have different  expectation values for a transformation with a short time segment (see Appendix A). 
This is the reason we introduced two moments $G_1$ and $G_2$.
In the same manner, we denote the moments of individual signal $v_j$ as 
\beq
D_1=\lla v_{j} v_{j}^*\rra,~~ D_2=\lla v_{j} v_{j}\rra,~~D_3=\lla (v_{j} v_{j}^*)^2\rra.
\eeq
Note that the above correlations can be estimated by taking the averages with respect to the dropped suffix $i$ for Fourier modes, as explicitly discussed in the next subsection ({\it e.g.} eqs.(\ref{q16}) and (\ref{q19})).

We can now write down  the correlation of two data $a\ne b$ as
\beq
\lla u_{a}u_{b}^*\rra=qD_1+G_1,~~\lla u_{a}u_{b}\rra=qD_2+G_2.\label{uu}
\eeq
Here, we used  the assumption that  detector noises $n_J$ are statistically independent, { and separate burst signals $v_i$ and $v_j$ ($i\ne j$) are uncorrelated.  }
In the same manner as the second-order moments,  the forth-order moment $\lla u_{a}u_{b}u_{c}^*u_{d}^*\rra$ is evaluated as
\beq
\lla u_{a}u_{b}u_{c}^*u_{d}^*\rra=qD_3+q(4D_1G_1+D_2G_2^*+D_2^*G_2)+q^2(2D_1^2+D_2D_2^*)+2G_1^2+G_2 G_2^*,\label{u4}
\eeq
and, using eqs.(\ref{uu}) and (\ref{u4}), we obtain
\beq
{\cal K}\equiv \lla u_{a}u_{b}u_{c}^*u_{d}^*\rra-\lla u_{a}u_{c}^*\rra\lla u_{b}u_{d}^*\rra -\lla u_{a}u_{b}\rra\lla u_{c}^*u_{d}^*\rra -\lla u_{a}u_{d}^*\rra\lla u_{b}u_{c}^*\rra =q\lla (v_{j} v_{j}^*)^2\rra=qD_3. \label{kurto}
\eeq
Note that this combination  is proportional to the averaged event number $q$ and does not have contribution from the cross terms. We can extract a quantity that is  purely arisen by discreetness of the underlying signals and  closely related to their  non-Gaussianity. Actually, the combination $\cal K$ is essentially  the same  as the Kurtosis parameter, a well known measure  in astrophysics to characterize non-Gaussianity (see {\it e.g.} \cite{Bernardeau:1993qu} for its application to the large-scale structure in the universe).  { For Gaussian variables $X_i$ ($i=1,2,3$ and 4), we have a simple identity $\lla X_1 X_2 X_3 X_4\rra=\lla X_1 X_2  \rra \lla X_3 X_4  \rra+\lla X_1 X_3  \rra \lla X_2 X_4  \rra+\lla X_1 X_4  \rra \lla X_2 X_3  \rra$, and we can easily confirm that the parameter ${\cal K}$ vanishes for Gaussian variables.  Indeed the right-hand-side of eq.(\ref{kurto}) does not depend on $G_1$ and $G_2$.}  A third-order moment called skewness is often used to characterize asymmetry of a probability distribution function for a scalar-type quantity around its mean.  However, since  GWs are   tensor quantities with no preferred signs, it is irrelevant to analyze the skewness parameter  here for characterizing non-Gaussianity of a GW background.

{ We should also comment on the relation between the parameter ${\cal K}$ and the central limit theorem.  Let us consider a situation where we increase the number of burst $q$, but fix the total power of the bursts $\sum_j^l v_j$ by $\lla (\sum_j^l v_j)(\sum_j^l v_j)^*\rra=qD_1=const$ and  also keep the ratio $D_3/D_1^2=const$ for the individual burst.   The parameter  ${\cal K}=q D_3=(D_3/D_1^2)(qD_1)^2 q^{-1}\propto q^{-1}$  asymptotically approaches to zero for  $q\to \infty$, and the data $(u_a,u_b,u_c,u_d)$ become more Gaussian-like. This matches with our expectation from the central limit theorem.}

\subsection{GW observation with multiple detectors}

In the previous subsection, we provide a simple demonstration for extracting  non-Gaussianity signature  caused by discreteness of burst signals.  Here, we specifically  discuss application of the  approach to a GW background.  As there are some gaps between the settings in the previous subsection and those in other parts of this paper, we do not straightforwardly  use the notations of variables defined in the previous subsection, but we introduce new (and mostly distinct) ones that will be used hereafter.

We assume to have multiple detectors (labels $J=a,b,\cdots$) that commonly have optimal sensitivity around a frequency $f$ with a bandwidth $\Delta f\sim f$, as expected for typical laser interferometers.  
We consider a signal analysis in the optimal band, and  neglect details of frequency dependence  ({\it e.g.} replacing an integral $\int (\cdots) df $ with a product  $(\cdots)_{f}\times \Delta f\sim (\cdots)_{f}\times f$).  In practice, this situation is approximately realized by applying a band-pass filter.

We model the data stream $s_J(t)$ of a detector $J$ in terms of a  GW signal $H_J$ and a detector noise $n_J$ as
\beq
s_J(t)=H_J(t)+n_J(t).
\eeq
For analyzing the latter $n_J$, it is advantageous to work in the Fourier space.   We decompose the data streams (total duration $T_{obs}$) into short segments of a given duration $T_{seg} (\gsim f^{-1}_{opt})$, and attach   a label $M(=1,\cdots,T_{obs}/T_{seg})$  for each segment.  Then we take Fourier transformations at discretized frequencies $f=N T_{seg}^{-1}$ ($N$: integer) as follows
%%%%%%%%%%%%%%%%%%%%%%%%%%%%%%%%%%%%%%%%%%%%%%%%%%%%%%%%%%%%%%%%%%%
\beq
s_{JM}(f)=\int_{(M-1)T_{seg}}^{MT_{seg}} e^{2\pi i f t}s_J(t)dt=H_{JM}(f)+n_{JM}(f),\label{fourier}
\eeq
with
%%%%%%%%%%%%%%%%%%%%%%%%%%%%%%%%%%%%%%%%%%%%%%%%%%%%%%%%%%%%%%%%%%%
\beq
H_{JM}(f)\equiv \int_{(M-1)T_{seg}}^{MT_{seg}}e^{2\pi i f t} H_J(t)dt,~~~
n_{JM}(f)\equiv \int_{(M-1)T_{seg}}^{MT_{seg}}e^{2\pi i f t} n_J(t)dt.
\eeq
%%%%%%%%%%%%%%%%%%%%%%%%%%%%%%%%%%%%%%%%%%%%%%%%%%%%%%%%%%%%%%%%%%%
The number of relevant Fourier modes in a segment is  $\sim  T_{seg}\Delta f \sim T_{seg} f$.

\if0
  In the Fourier transformation, we implicitly assumed to apply an  adequate time window function to suppress leakage of underlying frequency components to nearby modes due to  finiteness of the segment time $T_{seg}$.
\fi

We  assume that the detector noises $n_{JM}$ are stationary, Gaussian, independent, and have identical spectrum $S_N(f)$.  
Then the covariance  matrix for the detector noises is given by \footnote{In reality,  especially for short segment length $T_{seg}\Delta f\sim 1$, there is a weak correlation between nearby modes.}
%%%%%%%%%%%%%%%%%%%%%%%%%%%%%%%%%%%%%%%%%%%%%%%%%%%%%%%%%%%%%%%%%%%
\beq
\lla n_{JM}(f) n_{KL}(f')^* \rra\sim \frac12 \delta_{ML}\delta_{JK} \delta_{ff'} T_{seg} S_{N}(f). \label{spectral}
\eeq
%%%%%%%%%%%%%%%%%%%%%%%%%%%%%%%%%%%%%%%%%%%%%%%%%%%%%%%%%%%%%%%%%%%
Among the statistical assumptions about the detector noises,  independence between detectors  is the critical one for structure of our  approach, but other ones (including Gaussianity) would only modify statistical significance of data analysis \cite{Seto:2008xr}.

The total number of Fourier modes in observational time $T_{obs}$ is given by
\beq
N_t\sim\frac{T_{obs}}{T_{seg}}\times T_{seg} \Delta f=T_{obs}\Delta f .
\eeq
Since we will take  statistical averages of these Fourier modes with neglecting frequency dependence, we use a running index $i(=1,\cdots,N_t)$ for the modes without applying the double decomposition by $M$ and $f$ as in eq.(\ref{fourier}).
One of the reasons behind this prescription is that, as we see later, it is advantageous to take a short segment time $T_{seg}\sim (\Delta f)^{-1}$ for increasing the signal-to-noise ratio of the non-Gaussianity measurement. 

%But it is straightforward to rearrange our discussions for a longer segment length $T_{seg}$ from which we can take multiple ($\sim T_{seg}\Delta f)$ Fourier modes.

In the previous subsection, we find that the smooth Gaussian component $g$ in eq.(1) turned out to be irrelevant for the Kurtosis parameter.  Therefore, in the followings, we put aside such a contribution, and only consider the burst signals for GW signals
\beq
H_{Ji}=\sum_j^{l_i} R_{Jij},
\eeq
where $l_i$ is the total number of bursts in the time segment relevant for the Fourier mode $i$, and $R_{Jij}$ is response of a detector $J$ to a burst characterized by the labels $i$ (for Fourier modes) and $j$ (for bursts in a mode $i$).

We directly apply results in the previous subsection for our data now given by 
\beq
s_{ai}=H_{ai}+n_{ai},~~~s_{bi}=H_{bi}+n_{bi},~~~s_{ci}=H_{ci}+n_{ci},~~~s_{di}=H_{di}+n_{di}. \label{ndata}
\eeq
 We can basically identify the current data $s_{Ji}$ in eq.(\ref{ndata}) with the previous one $u_J$ in eq.(1) ($J$: the label for detectors). But, unlike the burst signals $u_j$ in eq.(1), we keep the label $J$  for the burst component $R_{Jij}$ to include geometrical dependence of a detector network that will be studied in the next section.
From  the data $s_{Ji}$ we take the  averages over the Fourier modes
\beq
C_{21ab}\equiv \frac1{N_t}\sum_i^{N_t} s_{ai} s_{bi}^*,~~~C_{22ab}\equiv \frac1{N_t}\sum_i^{N_t} s_{ai} s_{bi}. \label{q16}
\eeq
{ These contain statistical fluctuations around their expectation values.}
In the present setting, we have the expectation values
\beq
\lla C_{21ab}\rra = \lla s_{ai} s_{bi}^* \rra=\lla H_{ai}H_{bi}^*\rra= q\lla R_{a} R_{b}^* \rra,~~~
\lla C_{22ab}\rra = \lla s_{ai} s_{bi} \rra=\lla H_{ai}H_{bi}\rra= q\lla R_{a} R_{b} \rra\label{e15}
\eeq
with the average number of events $q=rT_{seg}$ ($r$: event rate) in a segment $T_{seg}$.
Here we used statistical independence of detector noises,  neglected frequency dependence, and also  omitted the label $j$ for bursts ({\it e.g.}  replacing $\lla R_{aij} R_{bij}^*\rra$ with $\lla R_{a} R_{b}^*\rra$).
We use the summations $C_{21ab}$ and $C_{22ab}$ as the estimators for the expectation values $\lla s_{ai} s_{bi}^*\rra$ and $\lla s_{ai} s_{bi}\rra$ as follows
\beq
C_{21ab}\to\lla s_{ai} s_{bi}^* \rra= q\lla R_{a} R_{b}^* \rra,~~~
C_{22ab}\to \lla s_{ai} s_{bi} \rra=q\lla R_{a} R_{b} \rra.
\eeq
The fluctuations of the summations $C_{21ab}$ and $C_{22ab}$ around their expectation values would be evaluated in subsection V.B.
Similarly, we define the summation
\beq
C_{4abcd}=\frac1{N_t}\sum_i^{N_t} (s_{ai} s_{bi}s_{ci}^* s_{di}^* )\ \label{q19}
\eeq
for the estimator of the forth-order moment $\lla (s_{ai} s_{bi}s_{ci}^* s_{ci}^* )\rra$. 
We then introduce the combination 
\beq
K=C_{4abcd}-C_{21ac}C_{21bd} -C_{21ad} C_{21bc}-C_{22ab} C_{22cd}^* \label{e18}
\eeq
as an estimator of the kurtosis corresponds to eq.(\ref{kurto})
\beq
\lla K\rra \to \lla s_{ai}s_{bi}s_{ci}^*s_{di}^*\rra-\lla s_{ai}s_{ci}^*\rra\lla s_{bi}s_{di}^*\rra -\lla s_{ai}s_{bi}\rra\lla s_{ci}^*s_{di}^*\rra -\lla s_{ai}s_{di}^*\rra\lla s_{bi}s_{ci}^*\rra =q \lla R_{a} R_{b} R_{c}^* R_{d}^* \rra.\label{e19}
\eeq
Therefore our observational target $K$ is characterized by a correlation of  four detectors $\lla R_{a} R_{b} R_{c}^* R_{d}^* \rra$ defined by an ensemble average of single burst event.

\section{formulation for generalized reduction functions}
In this section, we discuss dependence of the expectation value  $\lla R_a R_b R_c^* R_d^*\rra$  on the geometry of a detector network  and the polarization state of  incident  burst GWs.  Since the extra-Galactic  burst sources would  have random directions and orientations for realistic astrophysical models,  we need to deal with many averaging operations with respect to angular parameters of the detector-sources system.

\subsection{beam pattern functions  of detectors}
We  assume that the individual  burst duration $T_d$ is much shorter than  the time scale for transformation of the detector network  ({\it e.g.} $\sim$1 day for ground based detectors, $\sim$1 year for typical space interferometers). Thus we fix the network configuration  and introduce a fixed spherical coordinate system that is attached to the detector network ($D$-system, see figure 1).
We denote the source direction of the GW signal by  $\ven$ (equivalently, the propagation direction: $-\ven$) whose explicit form is given by \footnote{We attach a hat ${\hat{}}$ for an unit vector.}
\beq
\ven=(\cos{\phi_D} \sin{\theta_D},\sin{\phi_D}\sin{\theta_D},\cos{\theta_D}).
\eeq
Since we introduce many angular parameters for the geometry of the network and burst GW sources,    we also use the simplified notation "$D$" to represent the  two angular parameters $({\theta_D},{\phi_D})$ related to the  detector network.
In order to discuss GWs from the direction $\ven$, 
we define two unit vectors  $ {\hat \ve}_{\theta_D}$ and $ {\hat \ve}_{\phi_D}$ that are
normal to the source
direction $\ven$ and given  by
%%%%%%%%%%%%%%%%%%%%%%%%%%%%%%%%%%%%%%%%%%%%%%%%%%%%%%%%%%%%%%
\beq
{\hat \ve}_{\theta_D}=(\cos{\theta_D}\cos{\phi_D},\cos{\theta_D}\sin{\phi_D},-\sin{\theta_D}),~~
{\hat \ve}_{\phi_D}=(-\sin{\phi_D},\cos{\phi_D},0).
\eeq
%%%%%%%%%%%%%%%%%%%%%%%%%%%%%%%%%%%%%%%%%%%%%%%%%%%%%%%%%%%%%%
Then the  bases for transverse-traceless tensor $\ve^P$ $(P=+,\times)$ associated with the propagation direction $\ven$ 
are given as (see {\it e.g.} \cite{Flanagan:1993ix,allen}) 
%%%%%%%%%%%%%%%%%%%%%%%%%%%%%%%%%%%%%%%%%%%%%%%%%%%%%%%%%%%%%%
\beq
\ve^+_{}={\hat \ve}_{\theta_D} \otimes {\hat \ve}_{\theta_D}- {\hat \ve}_{\phi_D}
\otimes  {\hat \ve}_{\phi_D},
\quad
\ve^\times_{}={\hat \ve}_{\theta_D} \otimes
{\hat 
\ve}_{\phi_D}+{\hat  
\ve}_{\phi_D} \otimes {\hat \ve}_{\theta_D}. \label{pbase}
\eeq
%%%%%%%%%%%%%%%%%%%%%%%%%%%%%%%%%%%%%%%%%%%%%%%%%%%%%%%%%%%%%%

In this paper we only deal with simple L-shaped interferometers.  Response of such a detector 
 is characterized by a tensor $\ved_J$ given by  directions $\veu_J$ and $\vev_J$ of its two arms (see also \cite{Finn:2008vh}).  This tensor is traceless and written by
\beq
\ved_J=(\veu_J\otimes \veu_J-\vev_J\otimes \vev_J)/2.
\eeq
 The  beam pattern functions $F_J^P$ represent sensitivities of the detector to the two polarization modes defined in eq.(\ref{pbase}).  They  are  formally written with the tensor $\ved_J$ as
%%%%%%%%%%%%%%%%%%%%%%%%%%%%%%%%%%%%%%%%%%%%%%%%%%%%%%%%%%%%%%
\beq
F_J^P({\theta_D},{\phi_D})=\ved_J:\ve^P(\ven)=\sum_{ij}d_{Jij} e^P_{ij},
\eeq
with two polarization states $P=+,\times$. Note that $F_J^+$ has even parity, while $F_J^\times$ has odd parity.  These properties will  become useful for geometrical interpretation of our results derived later.

\subsection{response of detector to GWs}
Next we characterize GW emission from a burst source.  In addition to  the $D$ coordinate defined in previous subsection, we introduce another  coordinate system $S=(\theta_S,\phi_S)$ attached to the source (see figure 1). We  also define two polarization bases (plus ($p$) $\ve^p$ and  and cross ($c$) $\ve^c$  modes)\footnote{Here we use the labels $p$ and $c$ for plus and cross modes to distinguish the $+$ and $\times$ modes defined in the detector network coordinate.} associated with the source coordinate similar to eq.(\ref{pbase}). With these bases we decompose the polarization patterns of the burst GWs and represent their Fourier modes by $(h_p(f),h_c(f))$  that are defined at  the  origin of the detector network coordinate $D$.

Now we discuss the correspondence of two polarization bases $(\ve^+,\ve^\times)$ and $(\ve^p,\ve^c)$.
The polarization angle $\psi$ characterizes the rotation angle between two coordinate systems  $D$ and $S$  around the direction $\ven$ (see figure 1 and \cite{Thorne_K:1987}), and the decomposition  of the burst signals in the $D$ system $(h_+,h_\times)$ is given by   
\beq
h_+=(h_p \cos2\psi+h_c \sin2\psi),~~h_\times=(-h_p \sin2\psi+h_c \cos2\psi).\label{e25}
\eeq
Then the response of the detector $J$ to the incoming burst GW is given by
\beq
R_J=(F_J^+ h_+ +F_J^\times h_\times) \exp[-2\pi i f \vex_J\cdot\ven],\label{e26}\eeq
where we formally added the phase factor $\exp[-2\pi i f \vex_J\cdot\ven]$ induced by the position of the detector relative to the origin of the $D$ coordinate. { This prescription is valid, when the light-travel time between detectors are much smaller than the segment time $T_{seg}$ for Fourier transformation.}
With eqs.(\ref{e25}) and (\ref{e26}), the response $R_J$ is written as 
\beq
R_J={\cal F}_J^p h_p+{\cal F}_J^c h_c.
\eeq
The functions  ${\cal F}_J^{p,c}$ depend on the three angular parameters $D=({\theta_D},{\phi_D})$ and $\psi$.   The  information of the phase factor $\exp[-2\pi i f \vex_J\cdot\ven]$ is included in them as follows
\beqa
{\cal F}_J^p&=&(F_J^+ \cos(2\psi)+F_J^\times \sin(2\psi))\exp[-2\pi i f \vex_J\cdot\ven],\\
{\cal F}_J^c&=&(-F_J^+ \sin(2\psi)+F_J^\times \cos(2\psi))\exp[-2\pi i f \vex_J\cdot\ven].
\eeqa

As mentioned before, it is reasonable to assume that orientations and directions of the extra-Galactic  burst sources are randomly distributed. In other words, the source distribution is isotropic with  no global handedness. In this situation, we define the following three averaging operators for the  direction  angles $D=({\theta_D},{\phi_D})$, the orientation angles $S=(\theta_S,\phi_S)$ and  the polarization angle $\psi$; \footnote{ The angle $\psi$ is introduced to specify the orientation of the source frame relative to the detector frame. Since GWs are spin-2 quantities, we can limit the integral of $\psi$ in the range $[0,\pi]$, due to  the apparent identity between $\psi$ and $\psi+\pi$.}
\beq
[Z(D,S,\psi)]_{D}\equiv \frac1{4\pi}\int_0^{2\pi}  d{\phi_D} \int_0^\pi d{\theta_D} \sin{\theta_D}  Z(D,S,\psi),
\eeq
\beq
[Z(D,S,\psi)]_{S}\equiv \frac1{4\pi}\int_0^{2\pi}  d\phi_S \int_0^\pi d\theta_S\sin\theta_S Z(D,S,\psi),
\eeq
\beq
[Z(D,S,\psi)]_{\psi}\equiv \frac1{\pi}\int_0^{\pi}  d\psi  Z(D,S,\psi).
\eeq
We also use the simplified notations such as $[Z]_{DS}=[[Z]_D]_S=[[Z]_S]_D$.

%%%%%%%%%%%%%%%%%%%%%%%% Figure 1 %%%%%%%%%%%%%%%%%%%%%%%%%%%%
\begin{figure}
  \begin{center}
\epsfxsize=15cm
\epsffile{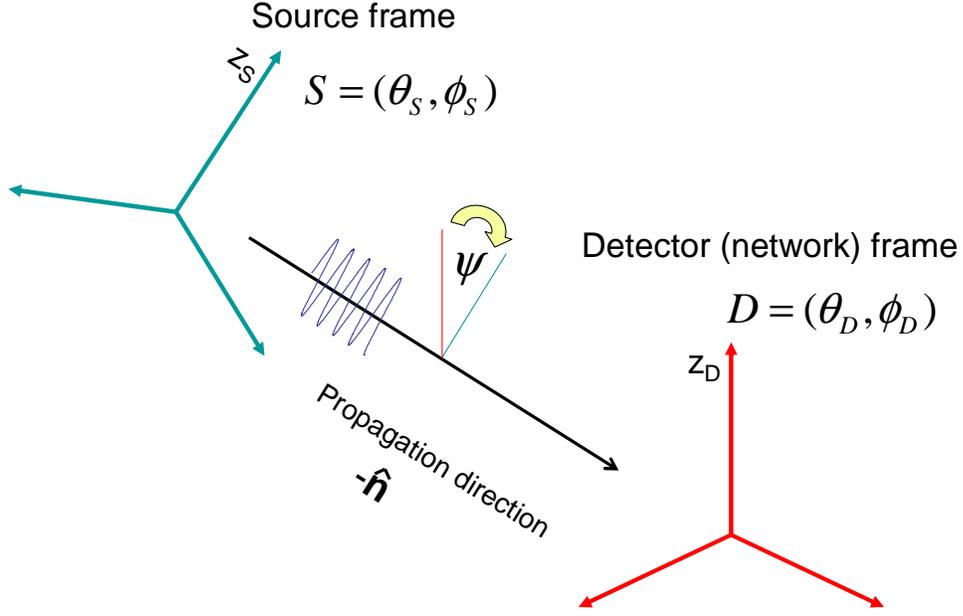}
 \end{center}

\vspace*{-0.5cm}

  \caption{In the detector network ($D$) frame, the direction of a GW source $\ven$ is characterized by two angles $D=({\theta_D},{\phi_D})$. In the source ($S$) frame, the direction of the detector network is given by $S=(\theta_S,\phi_S)$. On the plane normal to the propagation direction $-\ven$, projections of two $z$-axes ($\theta_D=0$ and $\theta_S=0$) are off-set by the polarization angle $\psi$.  These five angular parameters fix the relative configuration of two frames other than the distance between them.  
 }
\label{fig:f1}
\end{figure}
%%%%%%%%%%%%%%%%%%%%%%%%%%%%%%%%%%%%%%%%%%%%%%

\subsection{overlap reduction functions}
Here we calculate the geometrical averages $[R_a R_b^*]_{DS\psi}$ and $[R_a R_b  R_c^* R_d^*]_{DS\psi}$  to evaluate the expectation values   $ \lla R_a R_b^*\rra$ and $ \lla R_a R_b  R_c^* R_d^*\rra$ in eqs.(\ref{e15}) and (\ref{e19}).
For standard correlation analysis of a GW background, we take the combination $R_a R_b^*$ using two detectors $a$ and $b$. Its angular average is formally evaluated as \cite{Seto:2006hf,Seto:2007tn}
\beq
[R_a R_b^*]_{DS\psi }=\frac15 (\gamma_{Iab} [I]_S+\gamma_{Vab} [V]_S) \label{two},
\eeq
where  the Stokes parameters $I$ and $V$ are defined by \cite{radipro}
\beq
 I=|h_p|^2+ |h_c|^2,~~~V=i(h_ph_c^*-h_c h_p^*).
\eeq
The  parameter $I(\ge 0)$ represents the total intensity  of two polarization modes and has even parity.
The Stokes $V$ parameter is related to the circular polarization of waves and has odd parity. The $V$ parameter can be both positive and negative depending on the relative amplitude of right-handed waves ($h_R=(h_p+ i h_c)/{\sqrt 2}$) and left-handed waves ($h_L=(h_p- i h_c)/{\sqrt 2}$), and we have a relation $|V|\le I$ with equality only for 100\%-circularly polarized waves.
Note that these two real  parameters $I=|h_R|^2+|h_L|^2$ and $V=-|h_R|^2+|h_L|^2$  are invariant (spin-0) with respect to rotation  of two polarization bases around the propagation direction, while the combinations $h_R$ and $h_L$ themselves change as spin$\pm2$ quantities. Considering the fact that we take the averages for the  angle $\psi$ corresponding to the rotation, it is reasonable that our expression should be a simple combination of the spin-0 parameters $[I]_S$ and $[V]_S$.

  In this paper we consider GW sources with no preferred handedness, and  we have the identity $[V]_S=0$.  In other words, even if individual sources have circular polarization modes $V$ ({\it e.g.} inspiral binaries), its ensemble average should vanish.
In eq.(\ref{two}), the two functions $\gamma_I$ and $\gamma_V$ are overlap reduction functions defined by \cite{Flanagan:1993ix,allen,Seto:2007tn}
\beq
{\gamma_{Iab}}=\frac52[{\cal F}_a^p {\cal F}_b^{p*} +{\cal F}_a^c {\cal F}_b^{c*} ]_{D\psi}=\frac52[(F_a^+ F_b^++F_a^\times F_b^\times) e^{-2\pi i f (\vex_a-\vex_b)\cdot\ven/c}]_{D},\label{e35}
\eeq
\beq
{\gamma_{Vab}}=\frac{5}2 [-i({\cal F}_a^p {\cal F}_b^{c*} -{\cal F}_a^c {\cal F}_b^{p*} )]_{D\psi}=\frac52[-i(F_a^+ F_b^\times-F_a^\times F_b^+ )e^{-2\pi i f (\vex_a-\vex_b)\cdot\ven/c}]_{D}.\label{e36}
\eeq
These characterize correlated responses of two detectors $a$ and $b$ to incoming GWs for averaged source configurations.   To simplify our  notations, we hereafter omit  the subscript ${ab}$ for the overlap reduction functions.  With eq.(\ref{two}) and the identity $[V]_S=0$ for our  source model, the angular average of the product $R_a R_b^*$ is given by 
\beq
[R_a R_b^*]_{DS\psi }=\frac{\gamma_I}5 I_1,
\eeq
where we defined
\beq
I_1\equiv [I]_S.\label{ee40}
\eeq

So far, we have studied the averaging operations with respect to the geometrical parameters of sources relative to a  fixed detector network.  
 In addition to these geometrical ones, we need to deal with averaging for the intrinsic wave pattern of bursts, including  their  distance distribution. But they   are beyond  scope of this paper.  We rather assume that the parameter $I_1$ implicitly encompasses these additional averaging, and simply put
\beq
\lla R_a R_b^*\rra=\frac{\gamma_I}5 I_1.\label{e38}
\eeq

Now  we extend our analysis  to the four-point combination $\lla R_a R_b R_c^* R_d^*\rra$. 
After simple calculation, we find that the geometrical  average $[R_a R_b  R_c^* R_d^*]_{DS\psi }$ becomes a linear combination of the three parameters $[I^2]_S$, $[V^2]_S$ and $[IV]_S$ as follows
\beq
[R_a R_b  R_c^* R_d^*]_{DS\psi }=\frac{2}{25}\lmk \zeta_{II} [I^2]_S+\zeta_{VV} [V^2]_S+\zeta_{IV} [IV]_S \rmk \label{four}.
\eeq
Note that only the three parameters $I^2$, $V^2$ and $IV$ are the spin-0 quantities made from the relevant  fourth-order moments of  $h_p$ and $h_c$. In eq.(\ref{four}) we put the pre-factor 2/25 in order to simplify the expression given later in subsection V.B.
The coefficients $\zeta_{II}$, $\zeta_{IV}$ and $\zeta_{VV}$ are generalization of the overlap reduction functions for the four point correlations and defined by 
\beqa
\zeta_{II}&=&\frac{25}{16}[(3F_a^\times F_b^\times F_c^\times F_d^\times
+F_a^+F_b^+F_c^\times F_d^\times
+F_a^+F_b^\times F_c^+ F_d^\times
+F_a^\times F_b^+ F_c^+ F_d^\times 
+F_a^+F_b^\times F_c^\times F_d^+\nonumber \\
& & +F_a^\times F_b^+F_c^\times F_d^+
+F_a^\times F_b^\times F_c^+F_d^+
+3F_a^+F_b^+F_c^+F_d^+
)\exp[-2\pi i f (\vex_a+\vex_b-\vex_c-\vex_d)\cdot\ven]]_D,\label{e40}
\eeqa
%%%%%%%%%%%%%%%%%%%%%%%%%%%%%%%%%%%%%%%%
\beqa
\zeta_{VV}&=&-\frac{25}{16}[(F_a^\times F_b^\times F_c^\times F_d^\times
+3F_a^+F_b^+F_c^\times F_d^\times
-F_a^+F_b^\times F_c^+ F_d^\times
-F_a^\times F_b^+ F_c^+ F_d^\times 
-F_a^+F_b^\times F_c^\times F_d^+\nonumber \\
& & -F_a^\times F_b^+F_c^\times F_d^+
+3F_a^\times F_b^\times F_c^+F_d^+
+F_a^+F_b^+F_c^+F_d^+
)\exp[-2\pi i f (\vex_a+\vex_b-\vex_c-\vex_d)\cdot\ven]]_D,\label{e41}
\eeqa
%%%%%%%%%%%%%%%%%%%%%%%%%%%%%%%%%%%%%%%%%%
\beqa
\zeta_{IV}&=&\frac{25}8[(F_a^+ F_b^\times F_c^\times F_d^\times
+F_a^\times F_b^+F_c^\times F_d^\times
-F_a^\times F_b^\times F_c^+ F_d^\times
+F_a^+ F_b^+ F_c^+ F_d^\times 
-F_a^\times F_b^\times F_c^\times F_d^+ \nonumber \\
& & +F_a^+ F_b^+ F_c^\times F_d^+
-F_a^+ F_b^\times F_c^+F_d^+
-F_a^\times F_b^+F_c^+F_d^+
)\exp[-2\pi i f (\vex_a+\vex_b-\vex_c-\vex_d)\cdot\ven]]_D.
\eeqa
In this paper, we call them as {\it the generalized overlap reduction functions}.
Since we do not assume a global handedness, the average  $[IV]_S$ should vanish due to a parity reason.  But, here, it is important to note that our result in eq.(\ref{four}) depends on the quantity $[V^2]_S\ge 0$ that should not vanish for a burst model with $V\ne 0$ before the ensemble average. In contrast to $[IV]_S$ or   $[V]_S$, the cancellation between right- and left-handed modes does not occur for the parameter $[V^2]_S$. This, in principle, allows us to  statistically study the circular polarization state of the burst by studying the non-Gaussianity of their background.   As an example, we examined the ratio $[V^2]_S/[I^2]_S$ for a quadrupole-type emission pattern $(h_c,h_p)\propto [(1+\cos^2\theta_S)/2,\pm i \cos\theta_S]$, and   obtained the result $[V^2]_S/[I^2]_S=69/71$. 

Similar to  the definition of the parameter $I_1\equiv [I]_S$, we define the two parameters $I_2$ and $V_2$ by
\beq
I_2\equiv [I^2]_S,~~~V_2\equiv [V^2]_S,
\eeq
including implicit  averaging operations other than the geometrical ones.  Then the fourth correlation $\lla  R_a R_b R_c^* R_d^* \rra$ is written as
\beq
\lla  R_a R_b R_c^* R_d^* \rra=\frac2{25} (I_2\zeta_{II}+V_2\zeta_{VV}).\label{e44}
\eeq
To simplify some of expressions derived later, we also  introduce  a non-dimensional quantity $W$ of order unity by
\beq
W\equiv \frac{1}{I_1^2} \lmk I_2\zeta_{II}+V_2{\zeta_{VV}}   \rmk, \label{pw}
\eeq
and then we have
\beq
\lla  R_a R_b R_c^* R_d^* \rra=\frac2{25} I_1^2 W.
\eeq

\section{examples of the overlap reduction functions $\zeta_{II}$ and $\zeta_{VV}$}
Our fiducial target in this paper is a GW background made by a superposition of short GW bursts.  As we see in the next section, it is preferable to take  a short segment time $T_{seg}$ when  analyzing such a background with a Fourier transformation. In this case, we cannot naively apply the phase factor such as $\exp[-2\pi i f (\vex_a-\vex_b)\cdot \ven/c]$  to evaluate the overlap reduction functions. This is because the coherent phase structure is not clear-cut for a  Fourier transformation with a short time segment. Therefore we simplify our analysis below by working in  the long-wave limit (namely $f |(\vex_a-\vex_b)|/c\ll 1$), and neglect the phase factors.  Under this limit, the parity structure of the overlap reduction functions  $\gamma_I$, $\gamma_V$, $\zeta_{II}$ $\zeta_{VV}$ and $\zeta_{IV}$ in eqs.(\ref{e35})(\ref{e36})(\ref{e40}) and (\ref{e41})  become particularly simple from the basic parity correspondences $F^+_J\to F^+_J$ and $F^\times_J \to -F^\times_J$ for  the beam pattern functions.  As a result, we can easily show the identities $\gamma_V=\zeta_{IV}=0$ for the odd parity modes.  In addition, the generalized overlap reduction function $\zeta_{II}$  becomes invariant with respect to  replacement of the detector indexes, due to its apparent symmetry in eq.(\ref{e40}). For example, we have $[R_a R_b R_c^* R_d^*]_{DS\psi}=[R_a^* R_b R_c R_d^*]_{DS\psi}$.
In this section, with the long-wave approximation, we evaluate the generalized overlap reduction functions $\zeta_{II}$ and $\zeta_{VV}$ for specific networks of interferometers both in the space and on the Earth.

\subsection{BBO/DECIGO type detectors}
Our first example is the Big Bang Observer (BBO) \cite{bbo} and the Deci-hertz Interferometer Gravitational Wave Observatory (DECIGO) \cite{Seto:2001qf}.  They are  future plans of GW observation in  space with the optimal band around 0.1-1Hz.  One of their main goals is to directly study a  GW background from early universe by correlation analysis.   However,  it was pointed out that burst GWs from supernovae of population III stars might be a strong foreground for detecting a GW background from the early universe \cite{Buonanno:2004tp,sandick}.
Therefore, the  method proposed in this paper could be particularly useful for BBO/DECIGO, and might enable us to discriminate whether a detected background is consistent with a smooth ({\it i.e.} composed by many degree of freedom) Gaussian-like one. In this subsection, we briefly discuss the detector configurations of the proposed missions and evaluate their generalized overlap reduction functions. We will return to the astronomical aspects of the supernova background  in subsection V.C.

 With BBO (and similarly for DECIGO), two sets of equilateral-triangle-shaped system (like LISA \cite{lisa}) $Y_1$ and $Y_2$ would be configured to realize a good sensitivity to a background with a large overlap $\gamma_I\sim1$.  We can make two effective L-shaped detectors $(A_1,E_1)$ and $(A_2,E_2)$ from the units $Y_1$ and $Y_2$ respectively (see figure 2).  For these four effective detectors,  analysis under the low-frequency-approximation will be fairly well around their optimal frequencies $f=0.1$-$1$Hz.

  These four detectors are placed on the same plane (see figure 2), and responses to GWs can be regarded as identical for  $A_1- A_2$ and $E_1- E_2$ pairs ($\gamma_{IA_1A_2}=\gamma_{IE_1E_2}=1$) in the low frequency approximation.  Meanwhile, in geometrical point of view, the two-detector pair $A_1-E_1$ (same for $A_2-E_2$) is  misaligned by $45^\circ$ on the plane ($\gamma_{IA_1E_1}=\gamma_{IA_2E_2}=0$). As for the detector  noises of the four data,  we assume that their spectra are the same and their correlation is negligible.  This is because of the independence of the noises of two systems $Y_1$ and $Y_2$ and  the geometrical symmetry within each triangle \cite{Prince:2002hp}.\footnote{{ Roughly speaking, $(A_i,E_i)$ are linear combinations of three basic data $(x_i,y_i,z_i)$ obtained at three vertexes using adjacent arms.  They are written  as $A_i=\frac{x_i-y_i}{\sqrt 2}$ and $E_i=\frac{x_i+y_i-2z_i}{\sqrt 6}$. The noises $(n_{xi},n_{yi},n_{zi})$ of the basic three data  apparently have correlation. But,  if  they are completely symmetric (namely, $\lla n_{xi} n_{xi}^* \rra=\lla n_{yi}, n_{yi}^* \rra=\lla n_{zi} n_{zi}^* \rra$ and $\lla n_{xi} n_{yi}^* \rra=\lla n_{yi} n_{zi}^* \rra=\lla n_{zi} n_{xi}^* \rra$),  we have $\lla n_{Ai} n_{Ei}^*\rra=0$ due to symmetric cancellations \cite{Prince:2002hp}.} In actual observation we need to carefully study the potential effects caused by residual noise correlations.}  These  noise properties are highly preferable to measure the non-Gaussian parameter $K$ with the  four detectors $A_1$,  $A_2$, $E_1$ and $E_2$. 

Due to the geometrical simplicity of the network, we can analytically calculate the functions $\zeta_{II}$ and $\zeta_{VV}$, and  obtain 
\beq
\zeta_{II}=\frac5{14}. \label{bboi}
\eeq
 This result does not depend on the positions of two conjugates $*$ among the four data, as commented earlier.  On the other hand, we have the overlap functions for circular polarization modes as \footnote{ Here,  we can switch $A_1 \leftrightarrow A_2$ and $E_1 \leftrightarrow E_2$ for getting the same results.} 
\beq
\zeta_{VV}=-\frac{65}{63},~~~~(A_1A_2E_1^*E_2^*) \label{e48}
\eeq
 and 
\beq
\zeta_{VV}=\frac{85}{252}.~~~~(A_1A_2^*E_1E_2^*).\label{e49}
\eeq 
  Using the difference of the function $\zeta_{VV}$ as in eqs.(\ref{e48}) and (\ref{e49}), we can, in principle, measure the two parameters $I_2$ and $V_2$ separately from their linear combinations. The basic prescription for such separation was proposed in \cite{Seto:2006hf} to deal with the circular polarization mode in standard (two-point) correlation analysis (see also \cite{Nishizawa:2009bf}). 

%It is straightforward to apply a similar 
%method here (see also \cite{Nishizawa:2009bf}).

%%%%%%%%%%%%%%%%%%%%%%%% Figure 2 %%%%%%%%%%%%%%%%%%%%%%%%%%%%
\begin{figure}
  \begin{center}
\epsfxsize=15cm
\epsffile{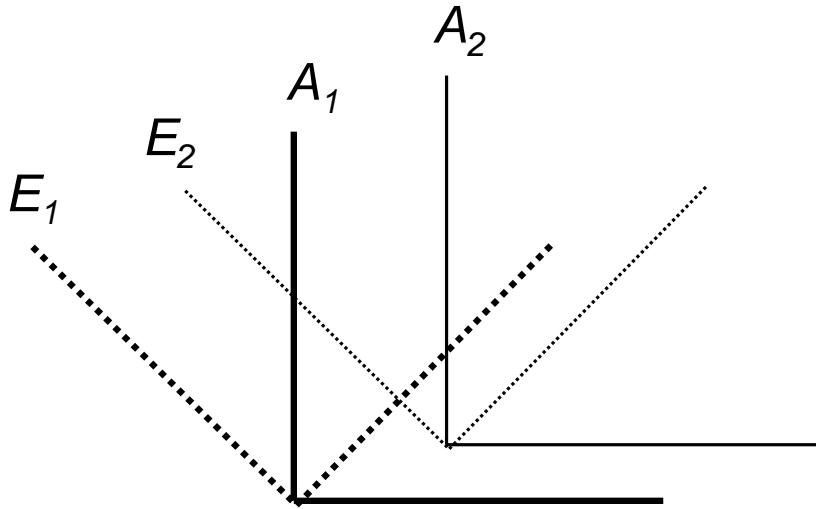}
 \end{center}

\vspace*{-0.5cm}

  \caption{Schematic picture for  the four  effective detectors $A_1$, $E_1$,  $A_2$ and $E_2$ made from two triangle units of BBO.  In the long-wave approximation, the positional differences between detectors can be neglected and responses to GWs can be identified for the $A_1-A_2$-pair and $E_1-E_2$-pair. Two detectors $A_1$$E_1$ has an off-set angle $45^\circ$ (same for $A_2E_2$), { and their noises are uncorrelated. }
 }
\label{fig:f2}
\end{figure}
%%%%%%%%%%%%%%%%%%%%%%%%%%%%%%%%%%%%%%%%%%%%%%

\subsection{Four detector network on the Earth}

We next evaluate the generalized overlap reduction functions $\zeta_{II}$ and $\zeta_{VV}$ for a detector network composed by the on-going and planned ground-based interferometers listed in Table 1. For detectors widely separated on the Earth, the characteristic frequency below which we can apply the low-frequency-approximation is given by $\sim c/(2\pi R_E)\sim 10$Hz ($R_E$: radius of the Earth $\sim 6000$km). This frequency is much smaller than the optimal frequency $\sim 100$Hz of the initial LIGO project and its advanced version \cite{adv}. Furthermore, the seismic noise level  would generally become significant  below $\sim 10$Hz for ground-based detectors.  Therefore, we cannot simply apply the low-frequency-approximation for a world-wide-detector network with their typical noise curves.  But it would be still interesting to examine the potential scientific advantage of fully using the four  detector network in the long-term run.  In addition our results might provide us with an useful insight for more realistic analysis.  In this spirit, we numerically evaluated the generalized overlap functions $\zeta_{II}$ and $\zeta_{VV}$.

We obtained $\zeta_{II}=0.07$, a relatively low value compared with the results  given in the previous subsection.   For the circular polarization mode,  our numerical results are 
  $\xi_{VV}=   -0.05$ for $CHL^*V^*$, 
   $ \xi_{VV}=0.011$  for $CH^*LV^*$  and 
  $ \xi_{VV}=-0.025$ for $CH^*L^*V$ with abbreviations $C$, $H$, $L$ and $V$ for the  four detectors given in Table 1.  The four detector network also has weak sensitivity to the circular polarization mode $V_2$.

%%%%%%%%%%%%%%%%%%%%%%%%%%%%%%%%%%%%%%%%%%%%%%%%%%%%%%%%%%%%%%%%%%%%%%%%%%
\begin{table}[!tb]
%\begin{ruledtabular}
\begin{tabular}{lccc}
detector & $\theta$ & $\phi$ & $\alpha$ \\
\hline
%\ AIGO ($A$) & 121.4 & 115.7 & -45.0\\
\ LCGT ($C$) & 53.6 & 137.3 & 70.0 \\
\ LIGO\ Hanford ($H$) & 43.5 & -119.4 & 171.8 \\
\ LIGO\ Livingston ($L$) & 59.4 & -90.8 & 243.0 \\
\ Virgo ($V$) & 46.4 & 10.5 & 116.5 
\end{tabular}
%\end{ruledtabular}
\caption{The position angles $(\theta,\phi)$ and the orientation angles $\alpha$ of the four ground-based 
 detectors (in units of degree).   The angles  $(\theta,\phi)$ are given for a spherical coordinate on the Earth with the north pole at $\theta=0^\circ$.  The orientation $\alpha$ is the angle between the local east direction and the bisecting line of two arms measured counter-clock wise. We use the abbreviations $C$, $H$, $L$ and $V$ as listed \cite{Abbott:2006zx,Kuroda:1999vi,Acernese:2002bw}.
\label{tab:detectors}}
\end{table}
%%%%%%%%%%%%%%%%%%%%%%%%%%%%%%%%%%%%%%%%%%%%%%%%%%%%%%%%%%%%%%%%%%%%%%%%%%

\section{signal-to-noise ratio of non-Gaussianity measurement}
In this subsection we derive expressions for the signal-to-noise ratio of  the non-Gaussianity measurement of a GW background made by superposition of burst signals.  First, in subsection V.A,  we write down the amplitude of the expectation values such as $\lla C_{21ab}\rra$ and $\lla K\rra$, including the finiteness of the time segment $T_{seg}$.  Then, in subsection V.B, we estimate their fluctuations due to detector noises.

\subsection{signal strength}

We represent the time profile of a burst GW in the source frame  by $(X_p(t),X_c(t))$, and take its Fourier transformation $(k_p(f),k_c(f))$ as follows
\beq
k_{p}(f)\equiv\int_{-\infty}^\infty dt X_{p}(t) \exp[2\pi i f t],~~~k_{c}(f)\equiv\int_{-\infty}^\infty dt X_{c}(t) \exp[2\pi i f t],
\eeq
where an infinite time segment is used for above integrals.
The expectation value for the total power of a  burst is given by 
\beq
P(f)=\lla [|k_p(f)|^2+|k_c(f)|^2]_S \rra.
\eeq
Our target background is a superposition of these bursts, and its 
spectrum is written by the burst rate $r$ and the power $P$  by (see {\it e.g.} \cite{Seto:2008xr})
\beq
S_{GW}(f)=\frac{rP}{8\pi}.\label{s55}
\eeq
{ This spectrum is defined per solid angle and per polarization mode \cite{Flanagan:1993ix,allen}.}
Here  the well known normalized spectrum $\Omega_{GW}(f)$  is related to the spectrum $S_{GW}(f)$ as
\beq
S_{GW}(f)=\frac{3H_0^2}{32\pi^3 Gf^3}\Omega_{GW}(f),
\eeq
with the Hubble parameter $H_0$.
In addition to the rate $r$ and the power $P$, the burst duration $T_d$ is another principle parameter for characterizing the bursts and their background.

Next we discuss   analysis of the background   using the  Fourier modes $(h_p,h_c)$ transformed in a finite time segment $T_{seg}$ as
\beq
h_{p}(f)\equiv \int_{T_{seg}} dt X_{p}(t) \exp[2\pi i f t],~~~h_{c}(f)\equiv\int_{T_{seg}}  dt X_{c}(t) \exp[2\pi i f t]. 
\eeq
In these expressions, we omitted the label $M$ for  representing the initial time of the support of the integrals (see eq.(\ref{fourier})).
  The expected number of bursts in the segment  is given by
\beq
q=rT_{seg}\max[1,T_{d}/T_{seg}].\label{e54}
\eeq
Here the second factor  in the right-hand-side is provided to include the effect that a single burst event is covered with multiple segments for $T_d>T_{seg}$. 
Meanwhile the amplitude $I_1=\lla |h_p(f)|^2+|h_c(f)|^2 \rra$ in eq.(\ref{ee40}) is given by
\beq
I_1\sim \lla |k_p(f)|^2+|k_c(f)]|^2 \rra\min[1,T_{seg}/T_{d}]= P\min[1,T_{seg}/T_{d}]\label{e55},
\eeq
where the second factor represents the dilution of power due to a  segment time $T_{seg}$ shorter than the signal duration $T_d$.
Using eqs.(\ref{e15})(\ref{e38})(\ref{e54}) and (\ref{e55}), we have
\beq
\lla C_{21ab}\rra =q\lla R_a R_b^*\rra=\frac{qI_1\gamma_I}5=\frac{8\pi S_{GW}T_{seg}\gamma_I}{5}.
 \label{meanc}
\eeq
Note that, with the spectrum $S_{GW}$  defined  in eq.(\ref{s55}),  the amplitude for the traditional two point correlation $\lla C_{21ab}\rra$ does not depend on the burst duration $T_{d}$.  Nevertheless, this is not true for the Kurtosis parameter, as we see below.
We can evaluate  the expectation value $\lla K\rra=q\lla R_a R_b R_c^* R_d^*\rra$ in the same manner, and the result is given by
\beq
\lla K\rra=\frac2{25}q I_1^2 W=\frac{16\pi}{25} S_{GW}PT_{seg} W \min[1,T_{seg}/T_d]. \label{ek}
\eeq
Here the parameter $W$ is defined in eq.(\ref{pw}), and in the present case, we have 
\beqa
W&=&\frac{\zeta_{II} I_2+\zeta_{VV} V_2}{I_1^2}=\frac{\zeta_{II} \lla [|h_p(f)|^2+|h_c(f)|^2]^2 \rra-\zeta_{VV} \lla [h_p(f) h_c(f)^*-h_c(f) h_p(f)^*]^2 \rra }{\lla [|h_p(f)|^2+|h_c(f)|^2] \rra^2}\\
&\simeq&\frac{\zeta_{II} \lla [|k_p(f)|^2+|k_c(f)|^2]^2 \rra-\zeta_{VV} \lla [k_p(f) k_c(f)^*-k_c(f) k_p(f)^*]^2 \rra }{\lla (|k_p(f)|^2+|k_c(f)|^2) \rra^2},
\eeqa
and it does not depend on $T_{seg}$. Here we used eq.(\ref{e55}) and the relations such as $I_2\simeq \lla [|k_p(f)|^2+|k_c(f)|^2]^2 \rra \lnk \min[1,T_{seg}/T_d]\rnk^2$.
The segment time $T_{seg}$ is an adjustable parameter for signal analysis.
 When we increase the  length $T_{seg}$, there is a transition point at $T_{seg}=T_d (\gsim (\Delta f)^{-1})$ where the signal $\lla K\rra$ starts to decrease due to dilution of the power (see figure 3). We will discuss implication of this fact at the end of the next subsection.

 From eqs.(\ref{s55}) and (\ref{ek}) we can derive a simple relation $\lla K\rra \propto r^{-1}S_{GW}^2 $ for  the event rate $r$ and  the background level $S_{GW}(f)$ (or $\Omega_{GW}$).  For a fixed amplitude $S_{GW}$, the signal $\lla K\rra$ becomes more Gaussian-like for a larger event rate $r$, as indicated by a smaller  $\lla K\rra$.  This is quite reasonable from the central limit theorem.

%%%%%%%%%%%%%%%%%%%%%%%% Figure 3 %%%%%%%%%%%%%%%%%%%%%%%%%%%%
\begin{figure}
  \begin{center}
\epsfxsize=15cm
\epsffile{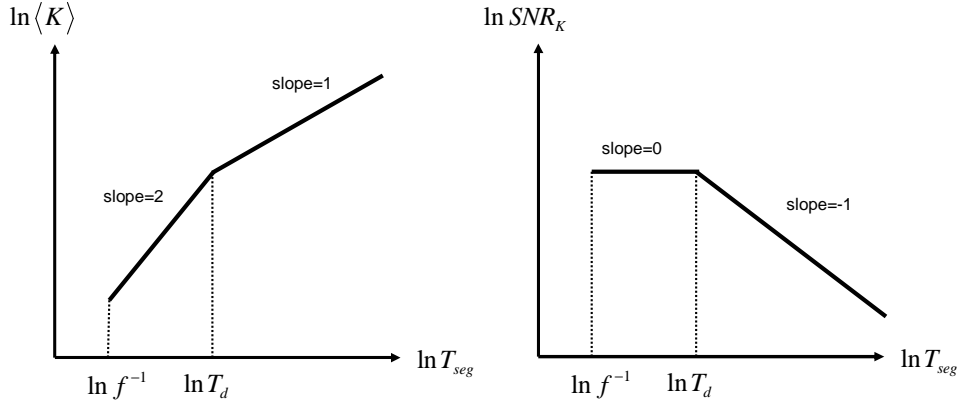}
 \end{center}

\vspace*{-0.5cm}

  \caption{Dependence of $\lla K\rra$ and $SNR_K$ on the adjustable parameter $T_{seg}$. The minimum value $f^{-1}$ of the  length $T_{seg}$ is set by the time resolution at the target frequency $f$ in interest. We can estimate the signal duration $T_d$ by identifying the transition of the slopes for the signal $\lla K\rra$.
 }
\label{fig:f3}
\end{figure}
%%%%%%%%%%%%%%%%%%%%%%%%%%%%%%%%%%%%%%%%%%%%%%

\subsection{RMS fluctuations by detector noises}

In this subsection we evaluate the expected noise level for  various statistical measures  such as $K$ or $C_{22ab}$.  
As a brief summary of notations, we begin our study with providing
the typical value of the optimal signal-to-noise ratio $SNR_{Bst}$ for a single burst
\beq
SNR_{bst}=\frac{2 (P\Delta f)^{1/2}}{5^{1/2}S_N^{1/2}}.\label{snrbst}
\eeq
Here the factor 2 originates  from the normalization  associated with definition of signal-to-noise ratio \cite{Thorne_K:1987} and the factor $1/5^{1/2}$ is due to  the  averaging operation for the angular responses of detectors.  Since we are interested in a GW background made by weak and undetectable bursts,  we assume $SNR_{Bst}\lsim 1$.

The correlation $C_{21ab}$ is given by a summation of the signal products $s_{ai}s_{bi}^*$ for two detectors $a$ and $b$ over the Fourier modes $i$ (see  eq.(\ref{q16})).
Hereafter  we assume that the GW spectrum $S_{GW}$ is smaller than the detector noise spectrum $S_N$ (weak signal condition), and therefore the fluctuations for our statistical measures are dominated by detector noises.  This corresponds to the condition when the correlation analysis becomes a powerful approach to detect a weak background buried among the detector noises, and is often assumed for theoretical analysis on GW backgrounds \cite{Flanagan:1993ix} (see also \cite{allen} for general cases).   From eqs.(\ref{s55}) and (\ref{snrbst}) we have $S_{GW}/S_N=\frac5{32\pi}SNR_{bst}^2 r/\Delta f$ and the condition for the assumption $S_{GW} \ll S_N$ is given by 
\beq
\frac5{32\pi}SNR_{bst}^2 \frac{r}{\Delta f} \ll 1. \label{cons}
\eeq

The root-mean-square (rms) fluctuation for a product $s_{ai}s_{bi}^*$ of each Fourier mode $i$ is given by  $S_N T_{seg}/(2\sqrt2)$  (see eqs.(\ref{fourier}) and (\ref{spectral})) with the  factor $1/\sqrt2$ for projection of the data toward the real axis on the complex plane \cite{Seto:2005qy}.
Then the rms fluctuation of the second-order moment  $\delta C_{21ab}$  is given by
\beq
\delta C_{21ab}\sim \frac{S_N  T_{seg}}{2\sqrt{2N_t}} \label{noise2},
\eeq
where  the  factor $1/\sqrt{N_t}$ ($N_t$: number of Fourier modes) is the statistical suppression of fluctuation due to the summation of independent $N_t$ modes. \footnote{There can be a weak correlation between noises of different Fourier modes, especially with a short segment length $T_{seg}$.  But we neglect it here.} 
We can derive the same result for the fluctuation $\delta C_{22ab}$ associated with the estimated moment $C_{22ab}$, and  put
\beq
\delta C_2=\delta C_{21ab}=\delta C_{22ab}.
\eeq

For two  aligned ($\gamma_{Iab}=1$) detectors,  the expectation value of $C_{21ab}$ becomes $\lla C_{21ab}\rra=\frac{8\pi S_{GW} T_{seg}}{5}$ (see eq.(\ref{meanc})), and   we obtain its signal-to-noise ratio  as
\beq
SNR_{C_{2}}\equiv \frac{\lla C_{21ab}\rra}{\delta C_2}=\frac{16\pi}5 \frac{S_{GW}}{S_N}(2 T_{obs}\Delta f)^{1/2}.\label{q65}
\eeq
The second power of this expression
 is essentially the same as the standard expression for correlation analysis in which we use the integral $\int df$ rather than the simple product  $\times \Delta f$ \cite{Flanagan:1993ix,allen}.

From eq.(\ref{e18}), the rms fluctuation  of the parameter $K$ due to the detector noises is estimated   as follows
\beq
\delta K\sim\max[\delta C_4, C_2\delta C_2, (\delta C_2)^2] \label{4noise}.
\eeq
Meanwhile the fluctuation for the term $C_4$ is evaluated as
\beq
\delta C_4\sim \frac{S_N^2T_{seg}^2}{4\sqrt{2N_t}}.
\eeq
The factors in this equation  can be  understood as in eq.(\ref{noise2}).
We can  evaluate the  ratios between the three elements in eq.(\ref{4noise}) \footnote{The second expression shows that we can neglect the bias induced by the nonlinear combination for $K$ in the weak signal condition.}
\beq
 \frac{C_2 \delta C_2}{\delta C_4}\sim\frac{16\pi}5\frac{S_{GW}}{S_N}<1,~~~\frac{(\delta C_2)^2}{\delta C_4}\sim \frac1{\sqrt{N_t}}\ll 1, \label{cc2}
\eeq
{ where we again assumed the weak signal case $S_{GW}\ll S_{N}$. If this condition does not hold, we need to deal with the contribution of the term $C_2 \delta C_2$.}
The relations (\ref{cc2}) show that, in the present setting,  the contribution from the term $C_4$ dominates the fluctuation for the parameter $K$.  
 Thus, for our weak signal case,  we have
\beq
\delta K\sim\delta C_4\sim \frac{S_N^2T_{seg}^2}{4\sqrt{2N_t}},
\eeq
and the signal-to-noise ratio for  the parameter $K$ is given by 
\beq
SNR_{K}=\frac{W \cdot SNR_{bst}^2 \cdot SNR_{C2}}{\Delta f \max[T_d,T_{seg}]}\label{snr4}.
\eeq
Here we used the relation 
$\min[A,B]=1/\max[A^{-1},B^{-1}]$.  In eq.(\ref{four}),  the normalization factor $2/25$ for the generalized overlap reduction  functions $\zeta_{II}$ and $\zeta_{VV}$  is determined in order to simplify the pre-factor in eq.(\ref{snr4}).  Since  the time resolution  at a frequency $ f$  effectively sets a limitation $ T_d\gsim f^{-1}\sim (\Delta f)^{-1}$, the minimum value of the denominator $\Delta f \max[T_d,T_{seg}]$ is $\sim 1$.

From eq.(\ref{snr4}), we can understand that, even if the individual burst signal is too weak ($SNR_{bst}\lsim1$) to be detected,  its basic properties can be statistically  studied with the amplification factor $SNR_{C2}$.  The factor $SNR_{C2}$ is the signal-to-noise ratio for standard correlation analysis (eq.(\ref{q65})) and proportional to $\Omega_{GW}T_{obs}^{1/2}$.  

{  In relation to this, we compare magnitudes of $SNR_{c2}$ and $SNR_K$.
Since we consider a background made by weak undetectable bursts $SNR_{bst}\lsim 1$ and also have relations $\delta f \max[T_d,T_{seg}]\gsim 1$ and $W=O(1)$, the signal-to-noise ratio $SNR_K$ would be generally smaller than that for the standard correlation analysis $SNR_{C2}$ (see also \cite{Drasco:2002yd} for a better performance of a non-Gaussian statistic).}

As a function of the adjustable parameter $T_{seg}(\gsim f^{-1})$, the signal-to-noise ratio $SNR_K$ is constant for a choice $T_{seg}\lsim T_d$, but it starts to decrease at   $T_{seg}\sim T_d$ (see figure 3).  The expectation value $\lla K\rra$ itself has a transition point at $T_{seg}\sim T_d$, as discussed after eq.(\ref{ek}). Therefore, once we can detect the non-Gaussianity parameter $ K$,  we can estimate the typical burst duration $T_d$   by analyzing its background and identifying  the transition.  The result for $SNR_K$ also shows that it is statistically preferable to take a  short segment time $T_{seg}$, as assumed in this paper (see figure 3).

In addition to the duration $T_d$, we might also estimate the rate $r$ and the characteristic power $P$ of weak bursts, if (i) they are assumed to be the dominant source of the total GW background at a band and (ii)  we  can simultaneously detect  signals  $C_2$ and $K$ for the background. These two signals would provide us with two combinations proportional to  $rP$ and $rP^2$.  Then we can separately obtain  the basic quantities $P$ and $r$ for the bursts   by putting $W\sim \zeta_{II}$. 

\subsection{GWs by Population III SNe}

In paper 1 we study a GW background caused by jet-like neutrino emissions at supernovae (SNe) of population III stars \cite{Buonanno:2004tp,sandick}. In the low frequency regime  at $f\lsim 1$Hz,  the individual burst wave-form can be regarded as a simple step-function like time profile known as {\it the burst  of memory} \cite{Epstein:1978dv}.  In relation to our analysis, its  characteristic signal duration is  $T_d\sim f^{-1}$ for a  Fourier mode at a  frequency $f$.  The amplitude of the circular polarization mode $V_2$ would be negligible, considering  the linear  emission pattern of neutrinos  with $I\gg |V|$ \cite{Epstein:1978dv}.   In this subsection, we reanalyze this background  with the expressions presented so far.

 The followings are the characteristic model parameters extracted from \cite{Buonanno:2004tp} and used in Paper 1;  the source redshift $z\sim 15$, the emitted neutrino energy $E_\nu\sim 10^{55}$erg, the mean anisotropy of the emission $\lla q\rra\sim 0.03$, and event rate $r\sim 0.01{\rm sec^{-1}}$.  But we should notice that the actual values of these parameters are quite uncertain, since our current knowledge on  the  population III SNe is highly limited.
For these model parameters, the amplitude  of the background spectrum is given by $\Omega_{GW}\sim 4\times 10^{-16}$ at $f\sim 0.3$Hz \cite{Seto:2008xr}.  With the scaling relation $\Omega_{GW}\propto rP$, we characterize the burst background using the combination of parameters $(\Omega_{GW},r)$ instead of the original ones $(P,r)$.

For the BBO noise spectrum in \cite{bbo}, the signal-to-noise ratio $SNR_{bst} \propto (\Omega_{GW}/r)^{1/2}$ of the individual burst is given by
\beq
SNR_{bst}\sim 0.6 \lmk \frac{\Omega_{GW}}{4\times 10^{-16}} \rmk^{1/2}   \lmk \frac{r}{0.01{\rm sec^{-1}}} \rmk^{-1/2} 
\eeq
for $f \sim 0.3$Hz.  { Note that the relation (\ref{cons}) holds for the typical model parameters described above (see also figure 4).}
We have the signal-to-noise ratio $SNR_{c2}$ for the correlation analysis with two overlapped detectors ({\it i.e.}  $\gamma_I=1$ as the $A_1$-$A_2$ pair in subsection IV) 
\beq 
SNR_{C2}\sim 80 \lmk \frac{\Omega_{GW}}{4\times 10^{-16}} \rmk  \lmk \frac{T_{obs}}{10{\rm yr}} \rmk^{-1/2} .
\eeq
From eq.(\ref{snr4})  the signal-to-noise ratio $SNR_K$ for the Kurtosis parameter $K$ becomes 
\beq
SNR_K\sim W~SNR_{bst}^2 ~SNR_{C2},
\eeq
for the optimal setting $T_{seg}\sim f^{-1}$ of the segment time.

With the generalized overlap reduction function $\zeta_{II}=5/14$ (see eq.(\ref{bboi})) for BBO  and the relation $W\sim \zeta_{II} I_2/I_1^2\sim \zeta_{II}$ \footnote{ Note that the ratio $I_2/I_1^2$ depends on the probability distribution of the burst amplitudes. The numerator $I_2$ is more affected by the stronger (but undetectable) ones. } for the non-dimensional parameter $W$ defined  in eq.(\ref{pw}),  we obtain the following result that is 
identical to eq.(18) in Paper 1 
\beq
SNR_K\sim \frac5{14} SNR_{bst}^2 ~SNR_{C2}\sim 10 \lmk \frac{\Omega_{GW}}{4\times 10^{-16}} \rmk^2  \lmk \frac{T_{obs}}{10{\rm yr}} \rmk^{-1/2} \lmk \frac{r}{0.01{\rm sec^{-1}}} \rmk^{-1}.
\eeq
This result  shows that, while the relevant astronomical parameters are highly uncertain,  the  non-Gaussianity signature $\lla K\rra$ might be detected for the population III SN background with BBO.   However, if the combination of the background parameters $\Omega_{GW}^2 r^{-1}$ is smaller than $\sim 2\times 10^{-29}$sec, the detection would be difficult. { In figure 4, we show the schematic picture of the signal-to-noise ratios $SNR_{C2}$ and $SNR_K$ as functions of $SNR_{bst}$ and $r$.}

%%%%%%%%%%%%%%%%%%%%%%%% Figure 4 %%%%%%%%%%%%%%%%%%%%%%%%%%%%
\begin{figure}
  \begin{center}
\epsfxsize=15cm
\epsffile{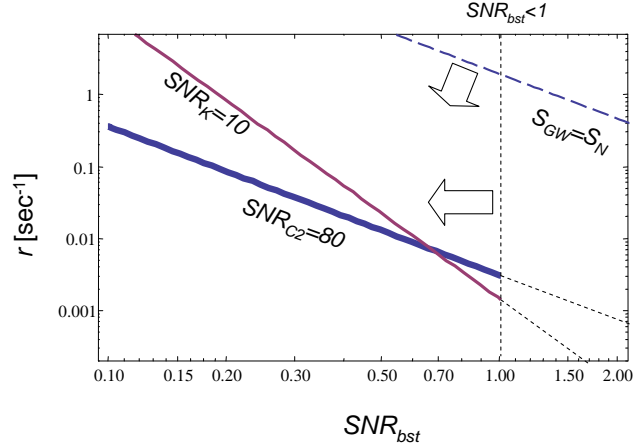}
 \end{center}

\vspace*{-0.5cm}

  \caption{Analysis for gravitational wave background from population III SNe with BBO. We have the scaling relations $SNR_{C2}\propto r SNR_{bst}^2$ and  $SNR_{K}\propto r SNR_{bst}^4$.
The typical  model parameters used in the main text are $SNR_{bst}=0.6$ (individual burst strength) and $r=0.01{\rm sec^{-1}}$ (burst rate). 
We put the observational time $T_{obs}=10$yr, the bandwidth $\Delta f$=0.3Hz and $W=5/14$.
The dashed line at the upper right is  $S_{GW}=S_{N}$ for the weak signal condition used in  our formulation (see  eq.(\ref{cons})). The vertical dashed line $SNR_{bst}=1$ shows the boundary for  weak undetectable signals.
 }
\label{fig:f4}
\end{figure}
%%%%%%%%%%%%%%%%%%%%%%%%%%%%%%%%%%%%%%%%%%%%%%

\section{summary and discussions}

In this paper we discussed an observational method to analyze a GW background made by a superposition of weak burst  signals that  are individually undetectable with small amplitudes.  Our  approach is to measure non-Gaussianity of the background induced by the discreteness of the bursts.     The  basic strategy was briefly introduced in  Paper 1.  This paper is a follow-up study with a significant enhancement on the basic formulation to clarify how we can study polarization properties of burst signals in response to the geometry of a detector network.

We find that our method enables us to measure linear combinations of two interesting parameters $I_2$ and $V_2$ defined by averaged squares of the Stokes parameters $I$ and $V$ of individual burst signals. These two parameters $I_2$ and $V_2$ can be separately estimated by devising data analysis with using multiple detectors, and they might provide us with an important insight to discriminate a GW background once detected by the standard correlation analysis with two detectors.

In this paper,  the shortness of the signal duration $T_d$ has been very important to show up the discrete nature of bursts by localizing their  power in a limited  time domain.  As an extension of our  method, it would  be  interesting to study non-Gaussianity induced by  almost monochromatic GW sources.  In this case the individual  signal is localized in the frequency domain, and it would be advantageous to use a relatively long time segment for Fourier transformation.

{ We have made several assumptions and simplifications on data analysis.  In order to apply the present method to real data of detectors, further careful works should be done. These include potential effects of non-Gaussianity and correlation of detector noises, and implementation on the short-term Fourier transformation. For example, if detectors have correlated non-Gaussian noises,  performance of our method would be limited.  In particular, two data streams  ($A_i,E_i$) obtained from one set of BBO (or DECIGO) could have correlated noises, depending on the symmetry of the three vertexes of the unit. This might reversely provide  interesting requirements on the mission designs of these future projects.
}

The author would like to thank T. Tanaka for helpful discussions. He also thanks anonymous referees for invaluable comments to improve the draft.
This work was 
supported by Grants-in-Aid for Scientific Research of the Japanese Ministry of Education, Culture, Sports, Science,
and Technology 20740151.

%\newpage
\appendix
\section{Fourier transformation with a finite time segment }

Here we summarize basic correlation properties of Fourier modes transformed  with a finite time segment $T_{seg}$.  We first express a time-domain signal $a(t)$ using the  continuous Fourier modes $A(f)$ (defined with  infinite time segment) as follows
\beq
a(t)=\int_{-\infty}^\infty e^{-2\pi i f t}A(f) df, \label{at}
\eeq
and 
\beq
A(f)=\int_{-\infty}^\infty e^{2\pi i f t}a(t) df.
\eeq
We assume that the signal $a(t)$ is a real function, or equivalently
$A(-f)=A(f)^*$. We define the power spectrum $S(f)$ of the modes $A(f)$ by
\beq
\lla A(f) A(f')^* \rra=\frac12 S(f) \delta (f-f').
\eeq

Next we evaluate the Fourier modes $\cA(f,T_{seg})$ defined with a finite segment length $T_{seg}$
\beq
\cA(f,{T_{seg}})=\int_0^{T_{seg}}e^{2\pi i f t } a(t) dt
\eeq
at discretized frequencies $f=N_t\times T_{seg}^{-1}$ ($N_t$: integer).  From eq.(\ref{at})
we can easily show the following relation;
\beq
\cA(f,{T_{seg}})=\int_0^{T_{seg}} e^{2\pi i f t }a(t) dt=\int_{-\infty}^\infty df' \frac{\sin(\pi (f-f'){T_{seg}})}{\pi (f-f')}A(f') e^{i\pi(f-f'){T_{seg}}}.\label{a3}
\eeq
Taking into account the finiteness of the segment $T_{seg}$ (see eq.(\ref{spectral})), the power spectrum $S(f,T_{seg})$ for the present case is given by 
\beq
S(f,{T_{seg}})=\frac2{T_{seg}} \lla\cA(f,{T_{seg}})\cA(f,{T_{seg}})^*    \rra=\int_{-\infty}^\infty df'   \frac{\sin^2\lkk \pi (f-f'){T_{seg}}\rkk}{\lkk \pi (f-f')\rkk ^2 {T_{seg}}} S(f') . \label{aa6}
\eeq
With the following asymptotic relation for the Dirac's delta function $\delta (x)$
\beq
 \lim_{{T_{seg}}\to \infty} \frac{\sin^2\lkk \pi x{T_{seg}}\rkk}{\lkk \pi x\rkk ^2 {T_{seg}}} =\delta(x),
\eeq
the estimated power spectrum $S(f,T_{seg})$ coincides with the original one $S(f)$ in the limit $T_{seg}\to \infty$ as
\beq
\lim_{{T_{seg}}\to \infty}S(f,{T_{seg}})=S(f).
\eeq
Eq.(\ref{aa6}) shows that,
for a finite $T_{seg}$, the spectrum $S(f,T_{seg})$ has a dominant contribution around the target frequency $f$ with a bandwidth $\Delta f\sim T^{-1}_{seg}$.

Now we evaluate the correlation $2\lla \cA(f,T_{seg})\cA(f,T_{seg}) \rra/T_{seg}$ corresponding to the moment $G_2$ in eq.(\ref{e5}). From eq.(\ref{a3}), we have
\beq
\frac2{T_{seg}} \lla\cA(f,{T_{seg}})\cA(f,{T_{seg}})    \rra= \frac2{T_{seg}} \int_{-\infty}^\infty df'   \frac{\sin[\pi (f-f'){T_{seg}}]\sin[\pi (f+f')]}{\pi^2 (f^2-f'^2)} S(f') e^{2\pi i f {T_{seg}}} .
\eeq
  In the limit $T_{seg}\to \infty$ we have
\beq
\lim_{{T_{seg}}\to \infty}\frac2{T_{seg}} \lla\cA(f,{T_{seg}})\cA(f,{T_{seg}})    \rra= \lim_{{T_{seg}}\to \infty}\frac2{T_{seg}} \int_{-\infty}^\infty df'   [\delta(f-f')+\delta(f+f')] S(f') e^{2\pi i f {T_{seg}}} = 0,\label{g2}
\eeq
and   the expectation value for the amplitudes of real and imaginary parts of the mode $\cA(f,T_{seg})$ are the same, as shown from the real part of eq.(\ref{g2}). In addition, the imaginary part of eq.(\ref{g2}) represents that the correlation of the two parts vanishes.
However, these  properties do not hold at a finite $T_{seg}$.  Therefore, in contrast to the limit $T_{seg}\to \infty$,  we need to keep  the correlation $ \lla\cA(f,{T_{seg}})\cA(f,{T_{seg}})    \rra$ for a finite $T_{seg}$.

\section{Analysis in Paper 1}
In this appendix, we comment on the two overlooked points in Paper 1.  They are caused by improper handling of  Fourier modes derived with a finite time segment $T_{seg}$, as discussed in Appendix A. 

In Paper 1, we consider the simple product $(s_{E_1} s_{A_2}^*)^2$ for analyzing non-Gaussianity of a burst GW background. Here  the data $s_{E1}=H_{E1}+n_{E1}$ and  $s_{A2}=H_{A2}+n_{A2}$ are made by GW signals $H_{E1}$ and $H_{A2}$ and detector noises $n_{E1}$ and $n_{A2}$ with the labels $A_1$ and $E_1$ for detectors defined in subsection IV.A.  But, we have the correlations  for GW signals $\lla H_{E_1}(f) H_{E_1}(f)\rra\ne 0$ and $\lla H_{A_2}(f)^* H_{A_2}(f)^*\rra\ne 0$ whose amplitudes can become large for a short segment $T_{seg}$.   We thus need to subtract the relevant terms in order to extract a non-Gaussianity signature, as outlined in section II.A.

Furthermore, with a finite $T_{seg}$ we also have the correlations of detector noises $\lla n_{E_1}(f)^* n_{E_1}(f)^*\rra\ne 0$ and $\lla n_{A_2}(f)^* n_{A_2}(f)^*\rra\ne 0$ for the simple product $(s_{E_1} s_{A_2}^*)^2$, and  the detector noises have a  contribution to the expectation value of the product  $(s_{E_1} s_{A_2}^*)^2$.  This qualitatively changes the statistical character of the present problem and  decreases the signal-to-noise ratio of the non-Gaussianity measurement, compared with the combination, such as $(s_{E_1}s_{E_2} s_{A_1}^* s_{A_2}^*)$ used  in  subsection IV.A.

\end{document}